\documentclass[fleqn,usenatbib]{mnras}

\usepackage[T1]{fontenc}
\usepackage{ae,aecompl}

\usepackage{graphicx}	% Including figure files
\usepackage{amsmath}	% Advanced maths commands
\usepackage{amssymb}	% Extra maths symbols

\usepackage{txfonts} %old version of newtxtext, should be loaded after amsmath

\usepackage{epstopdf}
\usepackage{journals}

\usepackage[none]{hyphenat}
\hypersetup{draft}

\title[Gravitational instability and SF in NGC~628]{Gravitational instability and star formation in NGC~628}
\author[A. A. Marchuk]
{A. A. Marchuk$^{1}$\thanks{E-mail:
a.marchuk@spbu.ru}\\
$^{1}$St. Petersburg State University,
Universitetskij pr.~28, 198504 St. Petersburg, Stary Peterhof, Russia\\
}

\date{Accepted XXX. Received YYY; in original form ZZZ}

\pubyear{2018}

\begin{document}
\label{firstpage}
\pagerange{\pageref{firstpage}--\pageref{lastpage}}
\maketitle

%%%%%%%%%%%%%%%%%%%%%%%%%%%%%%%%%%%%%%%%%%%%%%%%%%%%%%%%%%%%%%%%%%%%%%
\begin{abstract}
The gas-stars instability criterion for infinitesimally thin disc was applied to the galaxy NGC~628. 
Instead of using the azimuthally averaged profiles of data the maps of the gas surface densities (THINGS, HERACLES), of the velocity dispersions of stars (VENGA) and gas (THINGS) and of the surface brightness of the galaxy (S$^4$G) were analyzed. 
All these maps were collected for the same region with a noticeable star formation rate and were superimposed on each other. 
Using the data on the rotation curve values of $Q_\mathrm{eff}$ were calculated for each pixel in the image. The areas within the contours $Q_\mathrm{eff}<3$ were compared with the ongoing star formation regions ($\Sigma_\mathrm{SFR}> 0.007 \, M_{\sun}$\,yr$^{-1}$\,kpc$^{-2}$) and showed a good coincidence between them. The Romeo-Falstad disc instability diagnostics taking into account the thickness of the stellar and gas layers does not change the result. 
If the one-fluid instability criterion is used, the coincidence is worse.
The analysis was carried out for the area $r < $ 0.5$r_{25} $. \citet{Leroy_etal2008} using azimuthally averaged data obtained $Q_\mathrm{eff} \approx 3-4$ for this area of the disc, which makes it stable against non-axisymmetric perturbations and gas dissipation, and does not predict the location of star forming regions.
Since in the galaxies the distribution of hydrogen and the regions of star formation is often patchy, the relationship between gravitational instability and star formation should be sought using data maps rather than azimuthally averaged data.
\end{abstract}
%%%%%%%%%%%%%%%%%%%%%%%%%%%%%%%%%%%%%%%%%%%%%%%%%%%%%%%%%%%%%%%%%%%%%

%%%%%%%%%%%%%%%%%%%%%%%%%%%%%%%%%%%%%%%%%%%%%%%%%%%%%%%%%%%%%%%%%%%%%
\begin{keywords}
instabilities -- ISM: kinematics and dynamics -- galaxies: individual:
NGC~628 -- galaxies: ISM -- galaxies: kinematics and dynamics -- galaxies: structure.
\end{keywords}
%%%%%%%%%%%%%%%%%%%%%%%%%%%%%%%%%%%%%%%%%%%%%%%%%%%%%%%%%%%%%%%%%%%%%

%%%%%%%%%%%%%%%%%%%%%%%%%%%%%%%%%%%%%%%%%%%%%%%%%%%%%%%%%%%%%%%%%%%%%
\section{Introduction}
%%%%%%%%%%%%%%%%%%%%%%%%%%%%%%%%%%%%%%%%%%%%%%%%%%%%%%%%%%%%%%%%%%%%%

The disc gravitational instability is often thought to play an important role in
the large-scale star formation in galaxies. The theory of this subject is well developed 
\citep{Quirk1972,Jog_Solomon1984,Elmegreen1995,Efstathiou2000,Rafikov2001,Romeo_Wiegert2011,Romeo_Falstad2013} 
and the existence of the relationship between disc instability and star formation was demonstrated by observations \citep{Kennicutt1989,Hunter_etal1998,Boissier_etal2003,Leroy_etal2008,Romeo_Wiegert2011,Hunter_etal2013,Zheng_etal2013,Yim_etal2014,Romeo_Fathi2015,Romeo_Fathi2016,Romeo_Mogotsi2017,Hallenbeck_etal2016,Garg_Banerjee2017} and 
simulations (e.g. \citealp{Li_etal2005,Li_etal2006,Goldbaum_etal2015,Goldbaum_etal2016,Inoue_etal2016,Fiacconi_etal2017}).
Recent investigations (e.g., \citealt{Krumholz2012}) have shown that this link is rather indirect. The processes involved in the formation of stars are multiple, very complex and operate on atomic to galactic size scales 
(see e.g. \citealp{McKee_Ostriker2007}).
Even so, the regions, which are gravitationally  unstable  according to the simple one-fluid gravitational criterion by \citet{G_LB1965}, can be rather easily calculated and compared with the regions of the ongoing star formation
\citep{Kennicutt1989,Hunter_etal1998,Martin_Kennicutt2001}.

The predictions of gravitational instability diagnostics are not always accurate. 
For example, a half of galaxies in the sample by \citet{Martin_Kennicutt2001} demonstrates the perceptible star formation rate and stable gaseous discs. One reason is that stars affect the instability level and a gaseous disc should be considered with stellar one simultaneously, as was shown by \citet{Jog_Solomon1984}. Another reason lie in fact that real galaxies are not axisymmetric systems and this effect is very difficult to take into account in the instability analysis. Finally, most of the works mentioned examine one-dimensional data, which are placed along the spectrograph's long slit (e.g. \citealp{Marchuk_Sotnikova2018}) or more often azimuthally averaged (e.g. \citealp{Kennicutt1989,Hunter_etal1998,Boissier_etal2003,Leroy_etal2008}). As galaxies usually demonstrate non-axisymmetric or irregular structure and a small filling factor for gas, azimuthally averaged data can lead to large uncertainties. There are a small number of works which examine data maps, rather than one-dimensional data profiles \citep{Yang_etal2007, Elson_etal2012}, because applying gravitational instability criterion needs many independent types of observational data or various assumptions if data are missed.

Even in the era of massive Integral Field Unit (IFU) surveys galaxies with all available data of interest are rare, because these surveys either cover only central parts of galaxies or have low spatial resolution. NGC~628 is one of the few exceptions. It is close, face-one well-studied galaxy, which listed in many surveys. Instability analysis in \citet{Leroy_etal2008} (hereafter L08) shows that azimuthally averaged disc is marginally stable within $r < 0.5 r_\mathrm {25}$ and can be unstable at $r > 0.5 r_\mathrm {25}$. At the same time, a lot of star forming regions are located in central parts of the galaxy. Thus NGC 628 is a good candidate to apply reasonable two-component model and find how the model can predict the location of star forming area, which are scattered unevenly across the disc. Recently \citet{Dib2017} have done similar analysis for almost the same data cubes, but they applied a modified criterion using Larson scaling laws and focused on the star formation rate (SFR) prediction only. They also treated stars as an isothermal fluid and took only 90 data points with a lower resolution 750~pc.

In this work the relationship between two-component gravitational instability and large-scale star formation for NGC~628 using a small number of assumptions and a huge number of data points in the galaxy image is investigated. The performed analysis is similar to that in previous work \citet{Marchuk_Sotnikova2018}. In Section 2 the data sources and necessary formulas are presented along with short method description. Section 3 contains obtained results and discussion about them.

%%%%%%%%%%%%%%%%%%%%%%%%%%%%%%%%%%%%%%%%%%%%%%%%%%%%%%%%%%%%%%%%%%%%%
\section{DATA AND METHOD}
%%%%%%%%%%%%%%%%%%%%%%%%%%%%%%%%%%%%%%%%%%%%%%%%%%%%%%%%%%%%%%%%%%%%%

NGC~628 (M74) is an SA(s)c galaxy with two grand-design spiral arms. The galaxy is visible almost face-on under inclination $i=7\degr$.
\citet{Herrmann2008} compared different estimates of the distance and found that $D=8.6$~Mpc is the most accurate estimate (see figure 5 in \citealp{Herrmann2008} and more recent paper by \citealp{Kreckel2017}), while L08 used $D=7.3$~Mpc. In this work more precise 8.6~Mpc distance and scale $24\,\arcsec/\mathrm{kpc}$ are used. The galaxy visual size is $r_{25} = 294\arcsec$ according to L08. NGC~628 was studied intensively and included in more than 1200 works according to NASA ADS. A complete overview of other unmentioned properties may be found in \citet{Zou2011} and references therein.

Provided analysis is based on VENGA \citep{VENGA,VENGAXCO} data cubes and utilizes the same area sized $5.2\times1.7$~arcmin across the center. It extends to $1-2$ exponential stellar scales in every direction from the galaxy center assuming the exponential scale value equal $65\arcsec$ (L08). In paragraphs below every type of needed data is described and conversion formulas  are given. For comparison reasons data conversion methods in this work follow a scheme by L08 except the formulas for the velocity dispersion of stars.

%%%%%%%%%%%%%%%%%%%%%%%%%%%%%%%%%%%%%%%%%%%%%%%%%%%%%%%%%%%%%%%%%%%%%
\begin{figure}
\includegraphics[width=1\columnwidth]{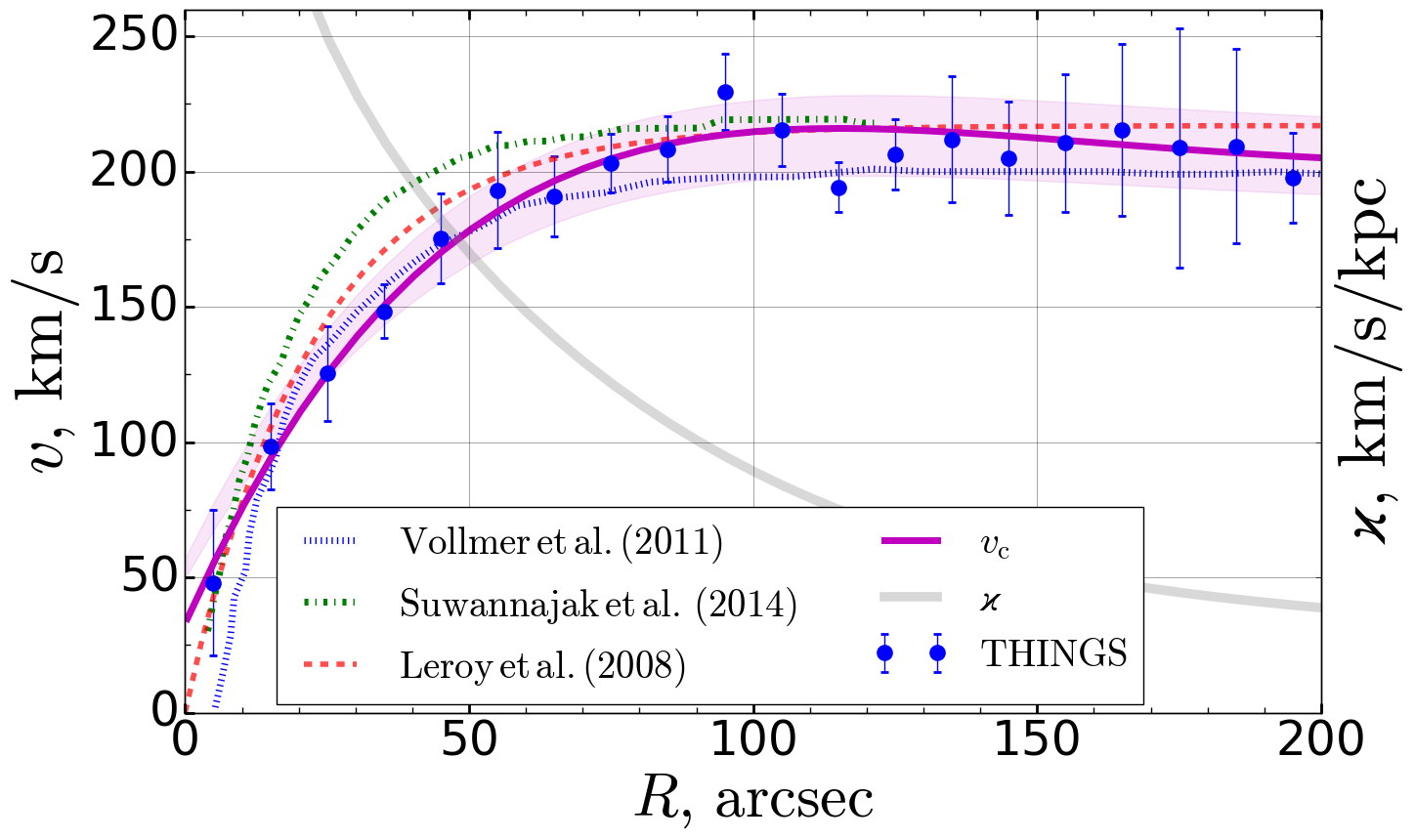}
\caption{Rotation curve. Points with errors represent THINGS velocity data, averaged over $10\arcsec$ bins. Solid magenta line shows gas rotation curve $\upsilon_\mathrm{c}$ fit. Solid grey line shows epicyclic frequency $\varkappa$. Other types of lines represent various approximations from the literature.}
\label{fig:rotvel}
\end{figure}
%%%%%%%%%%%%%%%%%%%%%%%%%%%%%%%%%%%%%%%%%%%%%%%%%%%%%%%%%%%%%%%%%%%%%

$\bf{HI\, data}$ Both atomic hydrogen surface densities $\Sigma_{\mathrm{H\, {\sc i}}}$ and velocity dispersions $\sigma_{\mathrm{H\, {\sc i}}}$ were obtained using THINGS \citep{THINGS} data cube. In this work the data cube with a natural weighting (NA) is used, which has size equal to $1024\times1024$~pixels and a spatial resolution equal to $1.5\arcsec$, or 62~pc. The integrated intensity was converted into $\Sigma_{\mathrm{H\, {\sc i}}}$ according to formula~(A1) from L08. The procedure includes helium correction factor 1.36. Obtained $\Sigma_{\mathrm{H\, {\sc i}}}$ and $\sigma_{\mathrm{H\, {\sc i}}}$ maps were compared with images in \citet{THINGS,Leroy_etal2008} and founded to be exactly the same. The maximal atomic hydrogen velocity dispersion value in the area of interest was around 23~km\,s$^{-1}$ and the mean value was close to 11~km\,s$^{-1}$, which L08 used as constant sound speed value for both atomic and molecular gases.

A first moment map from THINGS was also used to construct a ``cold'' gas rotation curve $\upsilon_\mathrm{c}$. Velocities along the major axis were extracted from the cube, bend across the center and corrected for the inclination and systemic velocity. The obtained profile was fitted by the cubic smoothing spline up to $200\arcsec$. The resulting $\upsilon_\mathrm{c}$ fit shown in Fig.~\ref{fig:rotvel} and is in good agreement with the results of other works by \citet{Suwannajak2014,Herrmann2009,Vollmer2011}, among which only the latter demonstrates the rotation curve which is long enough. 

The rotation curve was used to calculate the epicyclic frequency $\varkappa = \displaystyle \sqrt{2}\frac{\upsilon_\mathrm{c}}{R}\sqrt{1+\frac{R}{\upsilon_\mathrm{c}}\frac{d\upsilon_\mathrm{c}}{dR}}$, which is also presented in Fig.~\ref{fig:rotvel}. The derivative in this formula was calculated numerically and tested for stability by another fit using small degree polynomial and by using data from \citet{Vollmer2011}. As $\varkappa$, we actually use an azimuthally averaged characteristic. But there is no inconsistency in this approach. The epicyclic frequency is a derivative of the rotation curve, which, in turn, is a characteristic of the total potential. The potential is determined not by local fluctuations of the gas density, but by global massive subsystems. 

%%%%%%%%%%%%%%%%%%%%%%%%%%%%%%%%%%%%%%%%%%%%%%%%%%%%%%%%%%%%%%%%%%%%%
\begin{figure*}
\includegraphics[width=1.95\columnwidth]{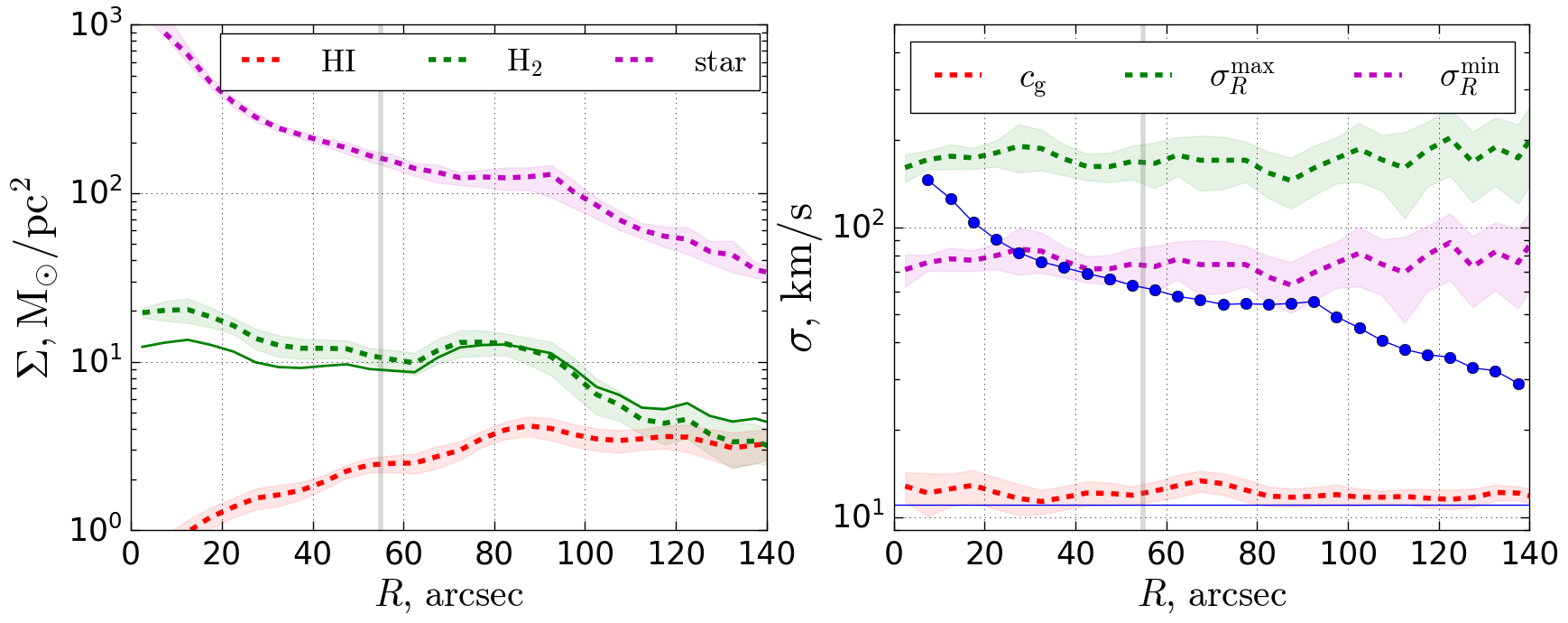}
\caption{Azimuthally averaged profiles for calculated data, where dashed lines represent mean values and filled areas show three standard deviations. Annuli are formed for $5\arcsec$ bins. Left plot shows surface densities $\Sigma$ and right plot shows velocity dispersions $\sigma$. Gaseous densities are not corrected for helium. Vertical grey line demonstrates radius of last full annulus ($55\arcsec$). Solid green line on the left shows change in mean $\Sigma_\mathrm{H_2}$ for another CO-to-$\mathrm{H_2}$ conversion factor (see text). Points on the right show $\sigma_R$ from L08 and horizontal line shows commonly accepted level $c_\mathrm{g}=11$~km\,s$^{-1}$.}
\label{fig:data}
\end{figure*}
%%%%%%%%%%%%%%%%%%%%%%%%%%%%%%%%%%%%%%%%%%%%%%%%%%%%%%%%%%%%%%%%%%%%%

$\bf{H_2\, data}$ Molecular gas surface densities were derived from HERACLES map \citep{HERACLES} in the line $\mathrm{CO}(J=2 \rightarrow 1)$. This map was obtained using beam size $13.6\arcsec$. Molecular gas follows spiral arms pattern (figure 17 in \citealp{HERACLES}). HERACLES map for NGC~628 demonstrates better sensitivity than data of BIMA-SONG \citep{BIMA2} and CARMA \citep{CARMA} surveys (see figure~1 in \citealp{VENGAXCO}). This figure also shows good agreement between all three surveys. The formula~(A3) from L08 was used to convert intensities into $\Sigma_\mathrm{H_2}$, where a constant CO-to-$\mathrm{H_2}$ factor $X_\mathrm{CO}=2\times10^{20}\, \mathrm{cm^{-2}(K\, km\, s^{-1})^{-1}}$ was assumed and resulting densities was corrected for helium. Map for part of analyzed region from \citet{Rebolledo2015} demonstrates values similar to those obtained here for $\mathrm{H_2}$. It should be noted that it is difficult to compare obtained map directly with central upper map in L08. While azimuthally averaged $\Sigma_\mathrm{H_2}$ profile is similar with that in L08 (see Table 7 in L08 for exact values), the same is not true for $\Sigma_\mathrm{H\, {\sc i}}$ profile. 
The source of this difference is unclear. It seems that on the figure~35 in L08 one profile was corrected for helium and another was not (see Fig.~\ref{fig:data} in comparison). In this work non-constant CO-to-$\mathrm{H_2}$ gradients from \citet{VENGAXCO} were also tested.

A molecular gas velocity dispersion map was found only in \citet{CO_disp} with inappropriate resolution. A comparison of radially averaged dispersions 
$\sigma_{\mathrm{H\, {\sc i}}}$ and $\sigma_\mathrm{CO}$ in \citet{Mogotsi_etal2016, CalduPrimo2013} suggests that both components are well mixed and justifies to use $\sigma_\mathrm{H_2}=\sigma_{\mathrm{H\, {\sc i}}}$. The colder molecular gas with $\sigma_{\mathrm{H\, {\sc i}}} = 1.5\times  \sigma_\mathrm{H_2}$ was also tested for a three-component model (see below).

$\bf{Stellar\, surface\, density}$ Stellar surface densities $\Sigma_\mathrm{s}$ were obtained using the same technique as in L08. They adopt a constant mass-to-light ratio in $K$-band equal to $0.5\:M_{\sun}/L_{\sun, K}$ and then apply an empirical conversion from $3.6\micron$ to $K$-band intensity using formula~(C1). L08 calculated stellar surface densities using data from SINGS \citep{SINGS}, while in this analysis the same formula was applied to more recent $3.6\micron$ data from S$^4$G \citep{Sheth2010}. A spatial resolution of data is near the same as in THINGS and is equal to $1.66\arcsec$. A resulting $\Sigma_\mathrm{s}$ map agrees well with a radial profile in L08 if azimuthal averaging is used (figure 35 in their work and Fig.~\ref{fig:data} here). Note that a more precise method for surface densities estimation from $3.6\micron$ (e.g. \citealp{Querejeta_etal2015}) has been introduced recently, but it is difficult to compare directly with previous results. 

$\bf{Stellar\, velocity\, dispersions}$ Reconstructing the stellar velocity dispersion in the radial direction $\sigma_R$ from line-of-sight observations is a difficult task \citep{Gerssen_etal1997,Marchuk_Sotnikova2017}. Three components of the stellar velocity ellipsoid (SVE) $\sigma_{\varphi}$, $\sigma_z$, $\sigma_R$ in azimuthal, vertical and radial directions respectively are connected with the line-of-sight velocity dispersion $\sigma_\mathrm{los}$ as follows 
%%%%%%%%%%%%%%%%%%%%%%%%%%%%%%%%%%%%%%%%%%%%%%%%%%%%%%%%%%%%%%%%%%%%%
\begin{equation}
\sigma^2_\mathrm{los} = \sigma_R^2\left[\left(\sin^2\phi + \frac{\sigma^2_{\varphi}}{\sigma_R^2}\cos^2\phi \right)\sin^2 i + \frac{\sigma_z^2}{\sigma_R^2}\cos^2 i\right],
\label{eq:sve}
\end{equation}
%%%%%%%%%%%%%%%%%%%%%%%%%%%%%%%%%%%%%%%%%%%%%%%%%%%%%%%%%%%%%%%%%%%%%
where $\phi$ is the angle between the slit and the major axis of the projected stellar disc. Instead of using common assumptions about the SVE form, some justified limitations were developed. The ratio $\sigma_\varphi / \sigma_R$ can be found from the equilibrium condition 
$\sigma_\varphi^2/\sigma_R^2 = 0.5\left(1 + \partial\ln \upsilon_\mathrm{c}/\partial\ln R\right)$ \citep{Binney_Tremaine2008}, where the derivative is calculated from the $\upsilon_\mathrm{c}$ fit. The rotation curve is smooth and the derivative is stable. The ratio $\sigma_z / \sigma_R$ was limited to $0.3 \le \sigma_z / \sigma_R \le 0.7$ as in \citet{Marchuk_Sotnikova2018}, where a lower boundary follows from the bending instability criterion \citep{Rodionov_Sotnikova2013} and the upper limit appears from observations \citep{Noordermeer_etal2008,Zasov_etal2008,Marchuk_Sotnikova2017,Pinna2018}. Substituting inequalities in Eq.~\eqref{eq:sve} gives the upper and the lower limits for $\sigma_R$ (hereafter $\sigma_R^\mathrm{max}$ and $\sigma_R^\mathrm{min}$) for every data point with known $\sigma_\mathrm{los}$. A data map of $\sigma_\mathrm{los}$ was taken from VENGA data cube (\citealp{VENGA}, see figure~16). The angle $\phi$ was measured from P.A.=$21\degr$. A comparison with L08 approach was made in Section~\ref{sec:discussion}.

$\bf{SFR}$ L08 suggested to use $FUV$ data along with $24\micron$ line intensity as SFR tracer. In this work corrected for dust an H$\alpha$ map from \citet{VENGAXCO} was used instead. This map has the same spatial resolution and size as VENGA data cube. Formula~(2) from \citet{Murphy2011} was used for the map conversion into SFR. \citet{VENGAXCO} also provides an SFR map constructed from $FUV$ and $24\micron$ data, which shows good agreement with the H$\alpha$ map, but has a slightly lower resolution. L08 also found that both methods coincide well.

%%%%%%%%%%%%%%%%%%%%%%%%%%%%%%%%%%%%%%%%%%%%%%%%%%%%%%%%%%%%%%%%%%%%
\begin{figure*}
\includegraphics[width=1.9\columnwidth]{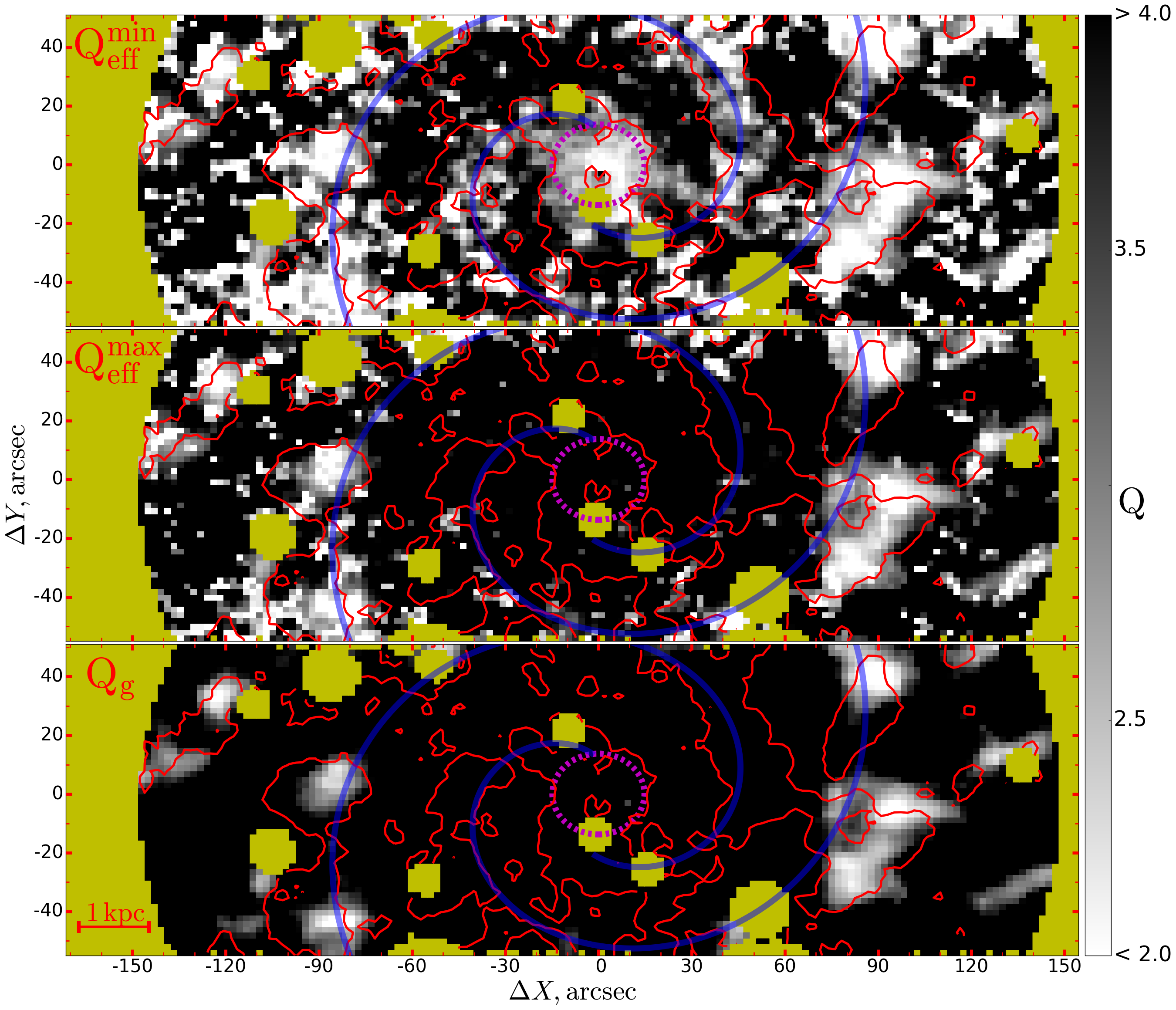}
\caption{$Q$-parameter for two-component with $\sigma_R^\mathrm{min}$ (upper), $\sigma_R^\mathrm{max}$ (middle) and gas-only (lower) models. Star forming regions $\Sigma_\mathrm{SFR}> 0.007 \, M_{\sun}$\,yr$^{-1}$\,kpc$^{-2}$ are shown by red contours. Masked stars are yellow. Bulge effective radius (magenta dashed) and logarithmic spirals (transparent blue, parameters from \citealp{Gusev2013}) are marked.}
\label{fig:fig1}
\end{figure*}
%%%%%%%%%%%%%%%%%%%%%%%%%%%%%%%%%%%%%%%%%%%%%%%%%%%%%%%%%%%%%%%%%%%%%

$\bf{Azimuthal\, profiles}$
Azimuthally averaged profiles for calculated data are shown in Fig.~\ref{fig:data}. The uncertainty in these profiles were calculated using formula (E1) from L08 and beam size for VENGA data cubes. It can be seen that relative scatter in surface densities is rather small in comparison with velocity dispersions. Note that $\Sigma$ profiles are in good agreement with L08 data except $\mathrm{HI}$ and that $\sigma_R$ velocity dispersions from L08 follow lower boundary of $\sigma_R^{\mathrm{min}}$ in this work. Notice also that all annuli for $R > 55\arcsec$ constructed not for the whole ellipses due to rectangular form of area under consideration. Probably this is the reason why scatter increased for large distances.

$\bf{Method}$
From all data cubes described above the same VENGA rectangle was extracted (see figure~1 in \citealp{VENGA}). Then coordinates of the center of $155\times55$ pixels grid in the VENGA cube were found and all needed data 
($\Sigma_{\mathrm{H\, {\sc i}}}$, $\sigma_{\mathrm{H\, {\sc i}}}$, 
$\Sigma_\mathrm{H_2}$, $\Sigma_\mathrm{s}$, $\sigma_{R}^\mathrm{max}$, $\sigma_{R}^\mathrm{min}$, $\Sigma_\mathrm{SFR}$, $\varkappa$) were collected for each pixel. In following instability analysis pixels masked as foreground stars in VENGA cube were not used. In order to avoid the bulge influence all points closer than the bulge effective radius $r_\mathrm{b,eff}=15\arcsec$ \citep{Mollenhoff_Heidt2001,Zou2011,Fisher_Drory2010}
were excluded from the analysis. Finally, outer regions $R > 140\arcsec$ of extracted area were also excluded because of significant uncertainties.

$\bf{Instability\, analysis}$ Two-component gravitational instability criterion was applied in the form derived by \citet{Rafikov2001}. One should found the 
dimensionless parameter $Q_\mathrm{eff}$ as
%%%%%%%%%%%%%%%%%%%%%%%%%%%%%%%%%%%%%%%%%%%%%%%%%%%%%%%%%%%%%%%%%%%%%
\begin{equation}
\frac{1}{Q_\mathrm{eff}} \equiv 
\max \left\{ 
\frac{2}{Q_\mathrm{s}} \frac{1}{\bar{k}}
\left[ 1-e^{-\bar{k}^2} I_{0}(\bar{k}^2) \right] + 
\frac{2}{Q_\mathrm{g}} s 
\frac{\bar{k}}{1 + \bar{k}^2 s^2} 
\right\} \, ,
\label{eq:rafikov}
\end{equation}
%%%%%%%%%%%%%%%%%%%%%%%%%%%%%%%%%%%%%%%%%%%%%%%%%%%%%%%%%%%%%%%%%%%%%
where $Q_\mathrm{g} \equiv \varkappa \, c_\mathrm{g} / \pi \, G \, \Sigma_\mathrm{g}$ and $Q_\mathrm{s} \equiv \varkappa \, \sigma_\mathrm{R} / \pi \, G \, \Sigma_\mathrm{s}$ are the dimensionless instability parameters for gaseous and stellar discs, $\Sigma_\mathrm{g} = 1.36\times(\Sigma_{\mathrm{H\, {\sc i}}} + \Sigma_\mathrm{H_2})$, $c_\mathrm{g} = \sigma_{\mathrm{H\, {\sc i}}}$ is the gas sound speed, $s \equiv {c_g}/{\sigma_R}$, $I_0$ is the modified Bessel function of the first kind and 
$\bar{k} \equiv k\,\sigma_R/\varkappa$ 
is a dimensionless wave number. Eq.~\eqref{eq:rafikov} is written for an infinitely thin disc. This assumption is discussed below. In simple axisymmetric case values of $Q_\mathrm{g}$ and $Q_\mathrm{eff}$ are compared with 1 and, if they are less, then a gaseous or a two-component disc respectively are considered unstable against gravitational perturbations. In more realistic nonaxisymmetric case the system became less stable and a comparison constant is equal to 2-3. The whole process and the background behind the two-component instability theory are described in detail in \citet{Marchuk_Sotnikova2018} and references inside.

%%%%%%%%%%%%%%%%%%%%%%%%%%%%%%%%%%%%%%%%%%%%%%%%%%%%%%%%%%%%%%%%%%%%%
\section{RESULTS AND DISCUSSION}
%%%%%%%%%%%%%%%%%%%%%%%%%%%%%%%%%%%%%%%%%%%%%%%%%%%%%%%%%%%%%%%%%%%%%
\label{sec:discussion}

Fig.~\ref{fig:fig1} demonstrates the results of gravitational instability diagnostics for NGC~628. The two-component model with smaller stellar radial dispersions $Q_\mathrm{eff}^\mathrm{min}$ shows a good agreement with observed star formation for a threshold $Q_\mathrm{eff}^\mathrm{min} < 2-3$ all over the area studied. On the other hand the parameter $Q_\mathrm{g}$ for gas only can explain star formation for distant areas $R > 80\arcsec$ but shows a stable disc inside this radius. The model with  $Q_\mathrm{eff}^\mathrm{max}$ represents an intermediate case between two mentioned. Unstable regions follow the spiral arms, which according to \citet{Foyle2010} just gather more gas inside and are not a trigger of star formation themselves. 
 
%%%%%%%%%%%%%%%%%%%%%%%%%%%%%%%%%%%%%%%%%%%%%%%%%%%%%%%%%%%%%%%%%%%%%
\begin{figure*}
\includegraphics[width=1.95\columnwidth]{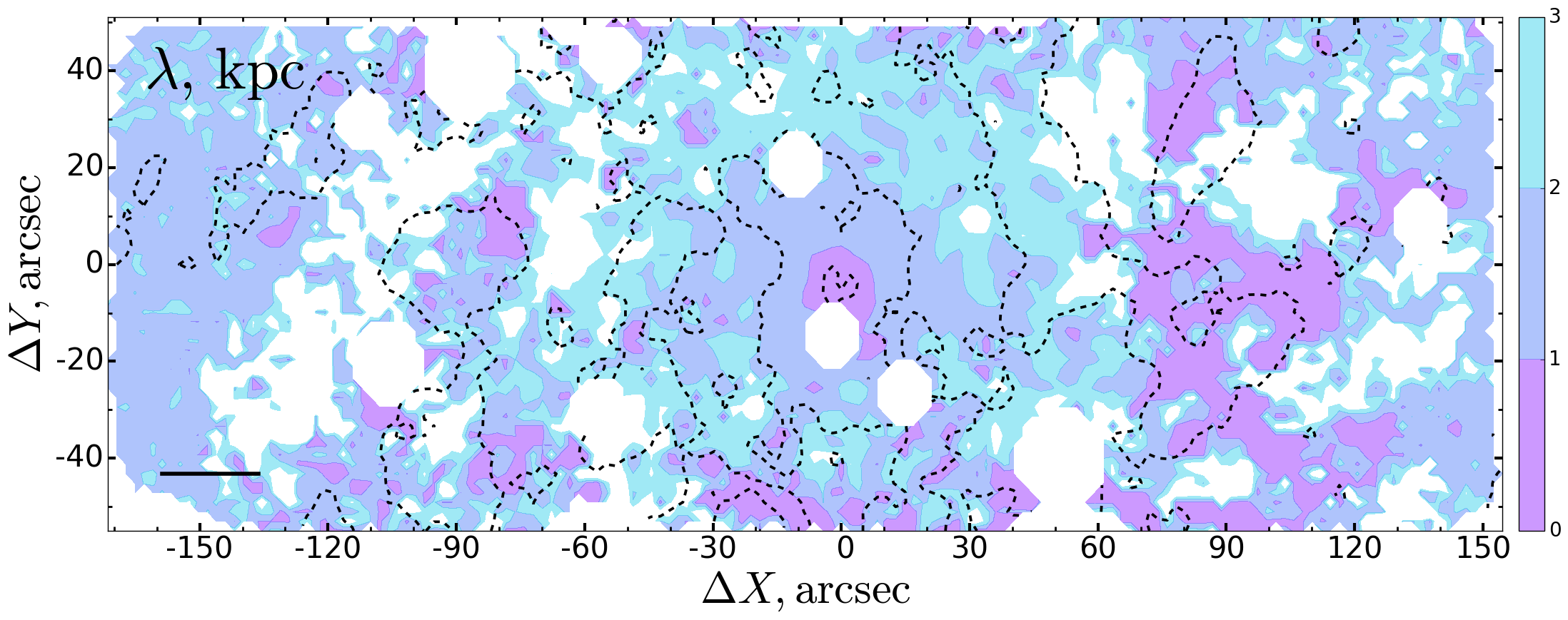}
\caption{The most unstable wavelengths $\lambda$ in kpc for $Q_\mathrm{eff}^\mathrm{min}$ model. Three different levels up to 3~kpc are shown. White areas demonstrate masked stars or regions with $\lambda > 3$~kpc. Black dashed lines represent areas of star formation. Black segment in the lower left corner shows 1~kpc linear scale.}
\label{fig:lambda}
\end{figure*}
%%%%%%%%%%%%%%%%%%%%%%%%%%%%%%%%%%%%%%%%%%%%%%%%%%%%%%%%%%%%%%%%%%%%%

\subsection{Uncertainties}

Correctness and stability of obtained results were tested in several ways. The analysis performed is correct only for a short-wavelength approximation, which should be checked. Even taking into account that some star forming areas show the most unstable wavelength $\lambda \approx 3$~kpc, corresponding regions always have the same or large length at least at one direction, as can be seen in Fig.~\ref{fig:lambda} for $Q_\mathrm{eff}^\mathrm{min}$ model. Secondly, uncertainties of $Q$ were checked. For $\mathrm{H}_2$ and $\sigma_\mathrm{los}$ error values are given in separated data cubes. For $\Sigma_{\mathrm{H\, {\sc i}}}$ the uncertainties were taken equal to 1~$M_{\sun}$\,pc$^{-2}$ as THINGS sensitivity limit. The rotation curve was varied within 7\% to account for differences between approaching and receding parts (transparent area in Fig.~\ref{fig:rotvel}). Stellar surface density uncertainties were obtained by varying the mass-to-light ratio for the $K$-band within 0.48-0.6~$M_{\sun}/L_{\sun}$ as L08 suggested. Applying all uncertainties together did not affect the results much and a root mean square error in $Q$ levels changes is around 15\%.

A common weakness in all such diagnostics is the adoption of a constant conversion factor $X_\mathrm{CO}$, since it can significantly change \citep{Oka1998,Pineda2010}. 
Fortunately, an attempt to measure the radial profile of $X_\mathrm{CO}$ was done in \citet{VENGAXCO} and \citet{Aniano2012}. 
The resulting stability levels were tested using the results by \citet{VENGAXCO}, as the authors mention large uncertainties in  \citet{Aniano2012} and caution against accepting a single $X_\mathrm{CO}$. Both provided profiles of $X_\mathrm{CO}$ for their parameters $N=1$ and $N=1.5$ almost coincided within the uncertainties and led to a change in the conversion factor of about 50\%, making it smaller in the center and larger outside (see solid line in Fig.~\ref{fig:data}, left subplot). 
Despite the noticeable changes in $\Sigma_\mathrm{H_2}$, taking into account the gradient $X_\mathrm{CO}$ does not significantly affect the values of $Q$, but slightly decreases it on the periphery. The next paragraph can shed light on this behaviour.

Disc thickness can significantly affect the disc stability and makes it more stable \citep{Romeo_Wiegert2011,Romeo_Falstad2013}. In order to account for this effect the approximation with individual components for atomic and molecular gas as in Eq.~(19) in \citet{Romeo_Falstad2013} was used. This approximation shows a good accuracy (see e.g. \citealp{Marchuk_Sotnikova2018}). Two cases $\sigma_{\mathrm{H\, {\sc i}}} = \sigma_\mathrm{H_2}$ and $\sigma_{\mathrm{H\, {\sc i}}} = 1.5\times\sigma_\mathrm{H_2}$ were tested. Comparison between $Q_\mathrm{RF}$ obtained for three-component thick model and exact $Q_\mathrm{eff}$ for a thin disc shows noticeable change. Taking thickness into account makes $Q_\mathrm{RF}^\mathrm{min}$ close to $Q_\mathrm{eff}^\mathrm{max}$ and $Q_\mathrm{RF}^\mathrm{max}$ becomes similar to $Q_\mathrm{g}$ for both gas dispersion suggestions. Fig.~\ref{fig:fig2} partially demonstrates this results, where $Q_\mathrm{RF}^\mathrm{min}$ is plotted. However the thickness stabilization effect can be compensated by the dissipation effect \citep{Elmegreen2011} up to its complete neglection in extreme cases. Unfortunately the dissipation effect is hard to account for. Note that used approximation from \citet{Romeo_Falstad2013} can also show which component is most unstable. In central parts of NGC~628 stars determine the dynamic status of a disc while molecular gas determines the instability for outer parts where $Q_\mathrm{g}$ model works well. This is the reason of result in previous paragraph, where change in CO-to-$\mathrm{H_2}$ factor does not affect central parts.

\subsection{Star formation relations}

%%%%%%%%%%%%%%%%%%%%%%%%%%%%%%%%%%%%%%%%%%%%%%%%%%%%%%%%%%%%%%%%%%%%%
\begin{figure*}
\includegraphics[width=1.95\columnwidth]{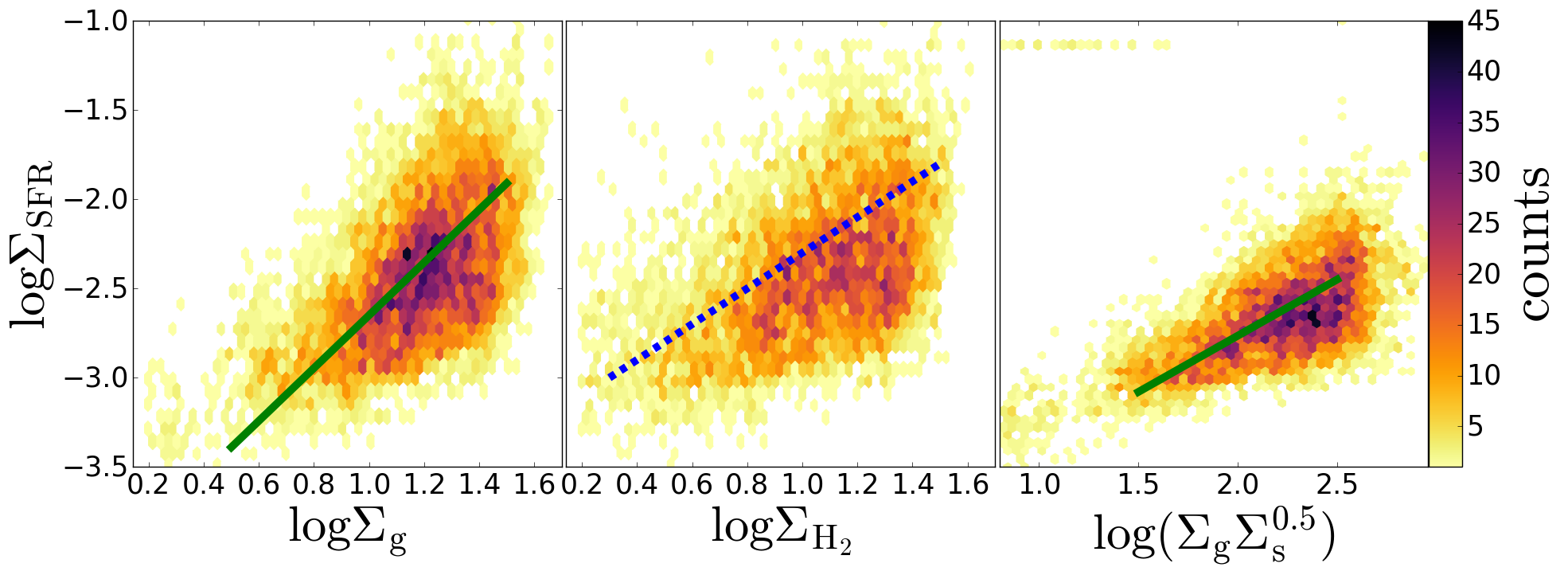}
\caption{Star formation laws. All plots represent point counts in hexagon binning with 50 bins. Left plot shows original Kennicutt-Schmidt law, line slope is 1.48. Central plot shows the same law for molecular gas only, dashed line represents constant depletion time $t_\mathrm{dep}$=2~Gyr. On the right image dependency proposed in  \citet{Ostriker2010} is shown, line slope is 1.}
\label{fig:ks}
\end{figure*}
%%%%%%%%%%%%%%%%%%%%%%%%%%%%%%%%%%%%%%%%%%%%%%%%%%%%%%%%%%%%%%%%%%%%%

%%%%%%%%%%%%%%%%%%%%%%%%%%%%%%%%%%%%%%%%%%%%%%%%%%%%%%%%%%%%%%%%%%%%%
\begin{figure*}
\includegraphics[width=1.95\columnwidth]{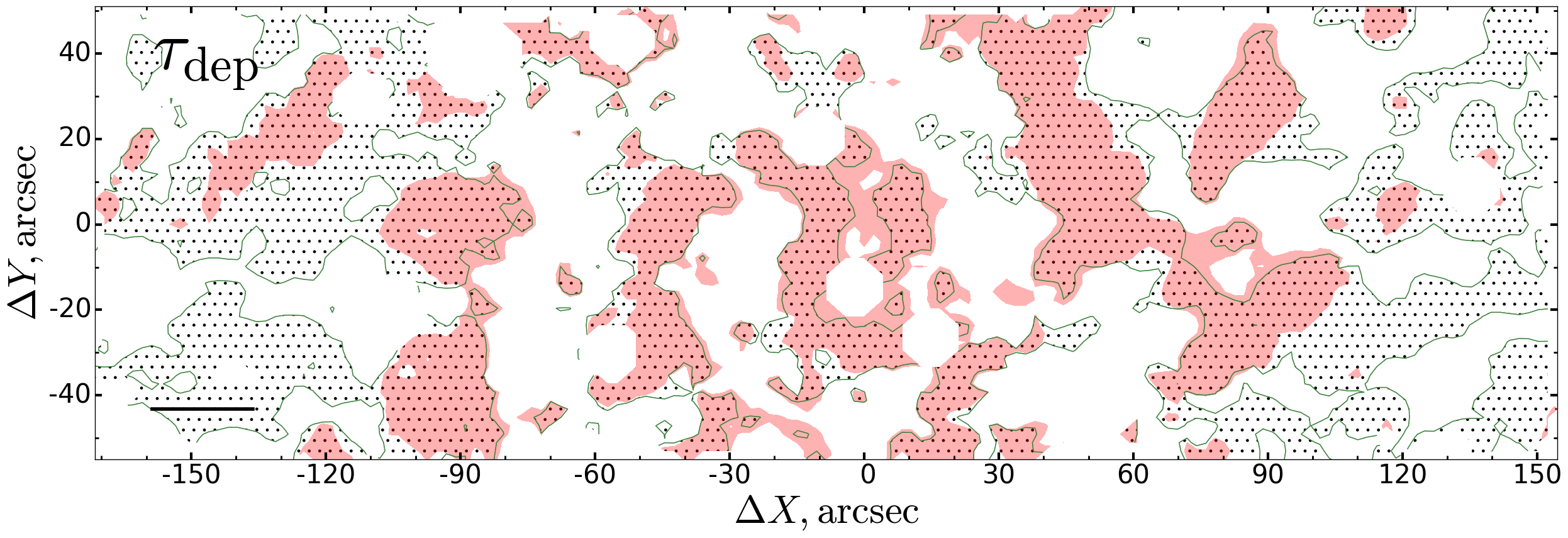}
\caption{Molecular gas depletion time in comparison with star-forming areas. Color filled areas correspond to the level $\Sigma_\mathrm{SFR}> 0.007 \, M_{\sun}$\,yr$^{-1}$\,kpc$^{-2}$. Hatched regions correspond to $t_\mathrm{dep}$<2~Gyr. Black segment in the lower left corner shows 1~kpc linear scale.}
\label{fig:depl}
\end{figure*}
%%%%%%%%%%%%%%%%%%%%%%%%%%%%%%%%%%%%%%%%%%%%%%%%%%%%%%%%%%%%%%%%%%%%%

Some well-known facts about star formation \citep{Kennicutt2012} were also tested. First one is the so-called Kennicutt-Schmidt star formation law \citep{Schmidt1959,Kennicutt1989}, which describes the relationship between SFR and the total gas density in $\Sigma_\mathrm{SFR} \propto \Sigma_\mathrm{g}^{N}$ form, where $N=1.4-1.5$ \citep{Kennicutt1989}. This and other relations are shown in Fig.~\ref{fig:ks}. Kennicutt-Schmidt law on the left subplot is fulfilled  with large scatter. It is probably show some evidences of more complex behaviour due to bended form. Best fit line is shown for $N=1.48$.

Central plot in Fig.~\ref{fig:ks} demonstrates an expected correlation between star formation and the amount of molecular gas. Values $\Sigma_\mathrm{H_2}$ on this plot are corrected for helium presence. Clearly Kennicutt-Schmidt law in NGC~628 is  determined by molecular gas mostly. Also expectation about no clear correlation between SFR density and atomic hydrogen is hold \citep{Bigiel2008}. No pixels was found with $\Sigma_{\mathrm{H\, {\sc i}}}$ higher than 10~$M_{\sun}$\,pc$^{-2}$, which is known as hydrogen molecularization threshold, as can be seen in Fig.~\ref{fig:data}.

Many other forms of Kennicutt-Schmidt law were developed in recent works. They are found from observations fits like $\Sigma_\mathrm{SFR} \propto \Sigma_\mathrm{g}^{1.13\pm0.05}\Sigma_\mathrm{s}^{0.36\pm0.04}$ \citep{Shi2011}, $\Sigma_\mathrm{SFR} \propto(\Sigma_\mathrm{s}^{0.5}\Sigma_\mathrm{g})^{1.09}$ \citep{Shi2018} as well as 
from sophisticated models like $\Sigma_\mathrm{SFR} \propto \Sigma_\mathrm{g} \Sigma_\mathrm{s}^{0.5}$ \citep{Ostriker2010} or $\Sigma_\mathrm{SFR} \propto \Sigma_\mathrm{g} \Sigma_\mathrm{s}$ (see references in \citealp{Krumholz2015}). The trend suggested by model in \citet{Ostriker2010}, where contribution to the vertical
gravitational field of the stellar disc plays important role in star formation, was checked. This law for analyzed data is represented on the right plot in Fig.~\ref{fig:ks}. It shows smaller scatter than original Kennicutt-Schmidt relation. The good confirmation of this trend also has been found recently in \citet{Westfall_etal2014}.

Finally, the map of molecular gas depletion time was studied. Many recent works (\citealp{Leroy2013} and references therein and in \citealp{Kennicutt2012}) have found that depletion time to be almost constant $t_\mathrm{dep}$=2~Gyr and this value already often has been  referenced as fiducial. In this work selected star formation areas (see below) are in excellent agreement with regions of constant depletion time in Fig.~\ref{fig:depl}. However, this is only allow to conclude that in selected areas of star formation all molecular gas can be  consumed faster than 2~Gyr if recent SFR persists. This constant level $t_\mathrm{dep}$=2~Gyr is also shown by line for central plot in Fig.~\ref{fig:ks}.

\subsection{Azimuthal averaging}

%%%%%%%%%%%%%%%%%%%%%%%%%%%%%%%%%%%%%%%%%%%%%%%%%%%%%%%%%%%%%%%%%%%%%
\begin{figure}
\includegraphics[width=0.95\columnwidth]{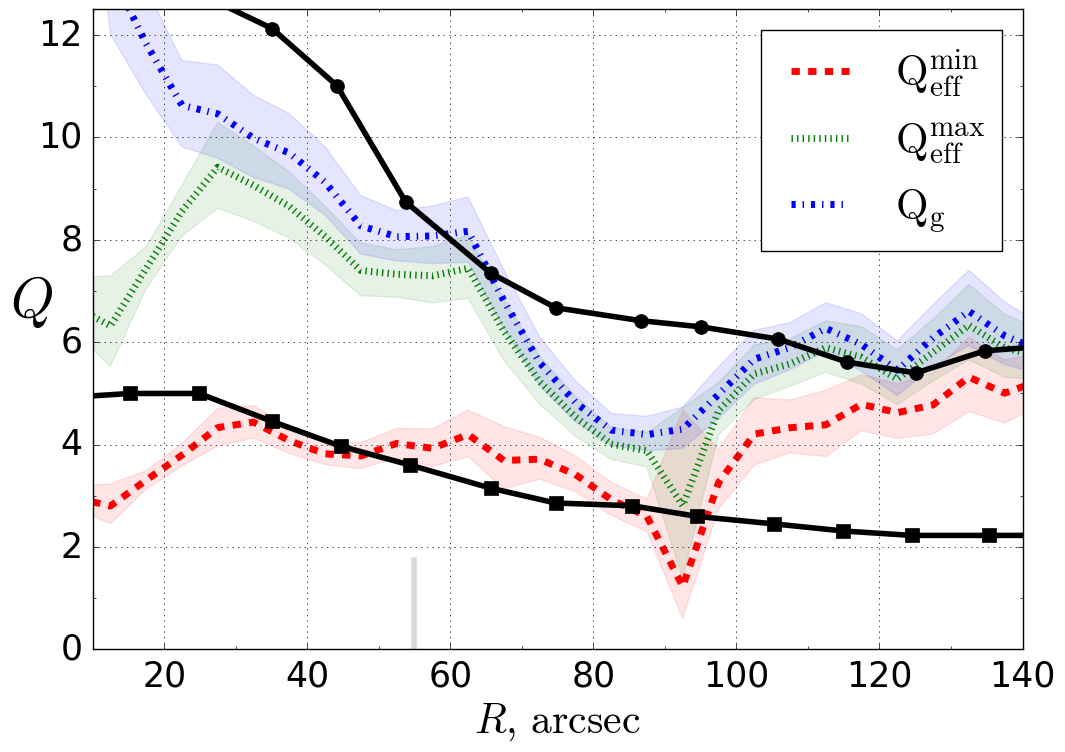}
\caption{Azimuthally averaged profiles for $Q$ data, where dashed lines represent mean values and filled areas show uncertainty, obtained for one standard deviation in data. Annuli are formed for $10\arcsec$ bins. Vertical grey segment on lower axis shows radius of last full annulus ($55\arcsec$). Solid black lines show $Q$ profiles from L08 for gas only (upper, circles) and for two-component (lower, squares) models.}
\label{fig:qaz}
\end{figure}
%%%%%%%%%%%%%%%%%%%%%%%%%%%%%%%%%%%%%%%%%%%%%%%%%%%%%%%%%%%%%%%%%%%%%

%%%%%%%%%%%%%%%%%%%%%%%%%%%%%%%%%%%%%%%%%%%%%%%%%%%%%%%%%%%%%%%%%%%%%
\begin{figure*}
\includegraphics[width=1.95\columnwidth]{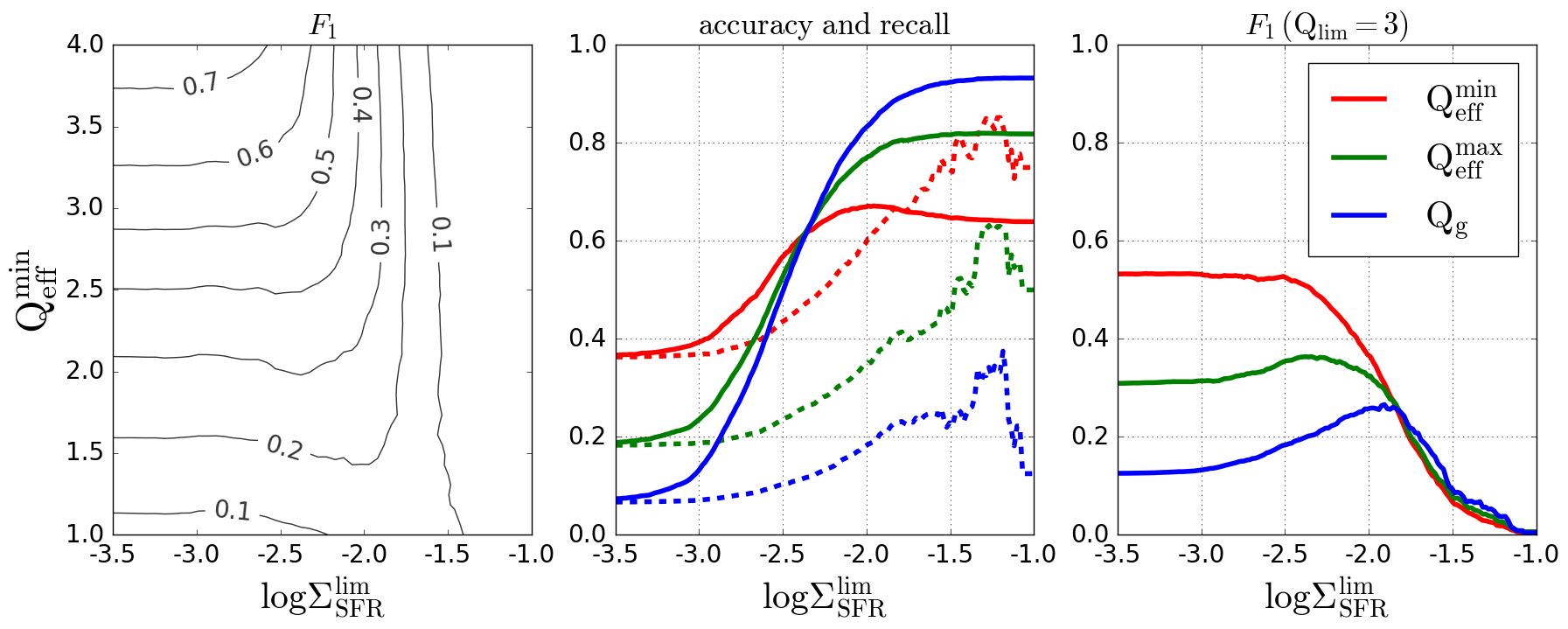}
\caption{Lines of the best coincidence between star formation and gravitational instability thresholds. On the left picture $F_1$ score contour map for $Q_\mathrm{eff}^\mathrm{min}$ model is shown. Central plot demonstrates accuracy (solid) and recall (dashed) scores for $Q_\mathrm{lim} = 3$ (see text for details). Colors are the same as for right plot. On the right plot $F_1$ scores with $Q_\mathrm{lim} = 3$ are presented.}
\label{fig:match}
\end{figure*}
%%%%%%%%%%%%%%%%%%%%%%%%%%%%%%%%%%%%%%%%%%%%%%%%%%%%%%%%%%%%%%%%%%%%%

All tests suggest that obtained two-component model is reliable and can associate the observed star formation regions with $Q \approx 2-3$ areas. However, this is not true if azimuthally averaged data are used. Azimuthal averaging is often used in works, because it has a physical justification that the whole disc is considered to be in an equilibrium state. Note, however, that the uneven distribution of gas has a little effect on the equilibrium, since its mass is relatively small. At the same time, regions with a larger gas density can be more unstable than others. Thus, spatial maps of the $Q$ stability parameter, although not fully consistent with the axisymmetric approximation, are a useful complement to radial profiles of $Q$ and can provide further insight between star formation and gravitational instability, for instance by revealing prominent local minima of $Q$ that would be suppressed by azimuthal averaging. In Fig.~\ref{fig:qaz} the same obtained results for models $Q_\mathrm{eff}^\mathrm{min}$, $Q_\mathrm{eff}^\mathrm{max}$ and $Q_\mathrm{g}$ are shown after azimuthal averaging procedure.  Notice how stable the discs become even within $55\arcsec$, where annuli were formed correctly. The uncertainties of $Q$ profiles are small and can not affect the results. The disc for all models in Fig.~\ref{fig:qaz} are stable, except central parts of $Q_\mathrm{eff}^\mathrm{min}$ profile and the location of massive bright spirals at $90\arcsec$. Such a stable disc does not allow predicting the location of star forming regions. 

These profiles are in agreement with those obtained in L08, but two-component $Q_\mathrm{eff}$ in L08 is systematically lower. The only major difference between diagnostics performed in this work and in L08 is different values of $\sigma_R$ used. In L08 authors did not use observational data, but instead use the assumption about
${\sigma_z}/{\sigma_R}=0.6$ and a constant stellar scale height. It led to 
%%%%%%%%%%%%%%%%%%%%%%%%%%%%%%%%%%%%%%%%%%%%%%%%%%%%%%%%%%%%%%%%%%%%%
\begin{equation}
\sigma_R = \frac{1}{0.6}\sqrt{\frac{2\pi G l_\mathrm{s}\Sigma_\mathrm{s}}{7.3}}, \label{eq:sigRh}
\end{equation}
%%%%%%%%%%%%%%%%%%%%%%%%%%%%%%%%%%%%%%%%%%%%%%%%%%%%%%%%%%%%%%%%%%%%%
where a factor 7.3 is justified by \citet{Kregel_etal2002} and $l_\mathrm{s}=2.3$~kpc was a stellar disc exponential scale length.
Obtained from Eq.~\ref{eq:sigRh} values of $\sigma_R$ are shown at right plot in Fig.~\ref{fig:data}. They are close to lowest boundary of $\sigma_R^\mathrm{min}$.  Applying such $\sigma_R$ values to the examined two-component model makes disc more unstable than the disc with $\sigma_R^\mathrm{min}$ or $\sigma_R^\mathrm{max}$ in this work. The same is also true in a less degree if one uses the isothermal velocity dispersion ellipsoid as was done by \citet{Dib2017}. Even for such lower dispersions profile of two-component gravitational instability criterion in L08 is still stay marginally stable $Q_\mathrm{eff} \approx 3-4$ in central regions and can not explain star formation at $r < $ 0.3 $r_{25}$. In addition notice that constant sound speed $c_\mathrm{g}=$11~km\,s$^{-1}$ in L08 can not change the stability status due to small variations in this parameter.

In summarize, discs appear to be more stable after averaging. Even for $Q_\mathrm{eff}^\mathrm{min}$ model mean values equal 3-4 almost everywhere except central region and spirals at $90\arcsec$. Such artificially cold  discs are not allowing to predict regions of star formation correctly and only usage of two-dimensional maps like in Fig.~\ref{fig:fig1} can do it. Conclusion that azimuthal averaging can strongly affect the obtained values of $Q$ should be fulfilled for galaxies similar to NGC~628, where the matter distribution is patchy.

\subsection{Best match with SFR threshold}

%%%%%%%%%%%%%%%%%%%%%%%%%%%%%%%%%%%%%%%%%%%%%%%%%%%%%%%%%%%%%%%%%%%%%
\begin{figure*}
\includegraphics[width=1.9\columnwidth]{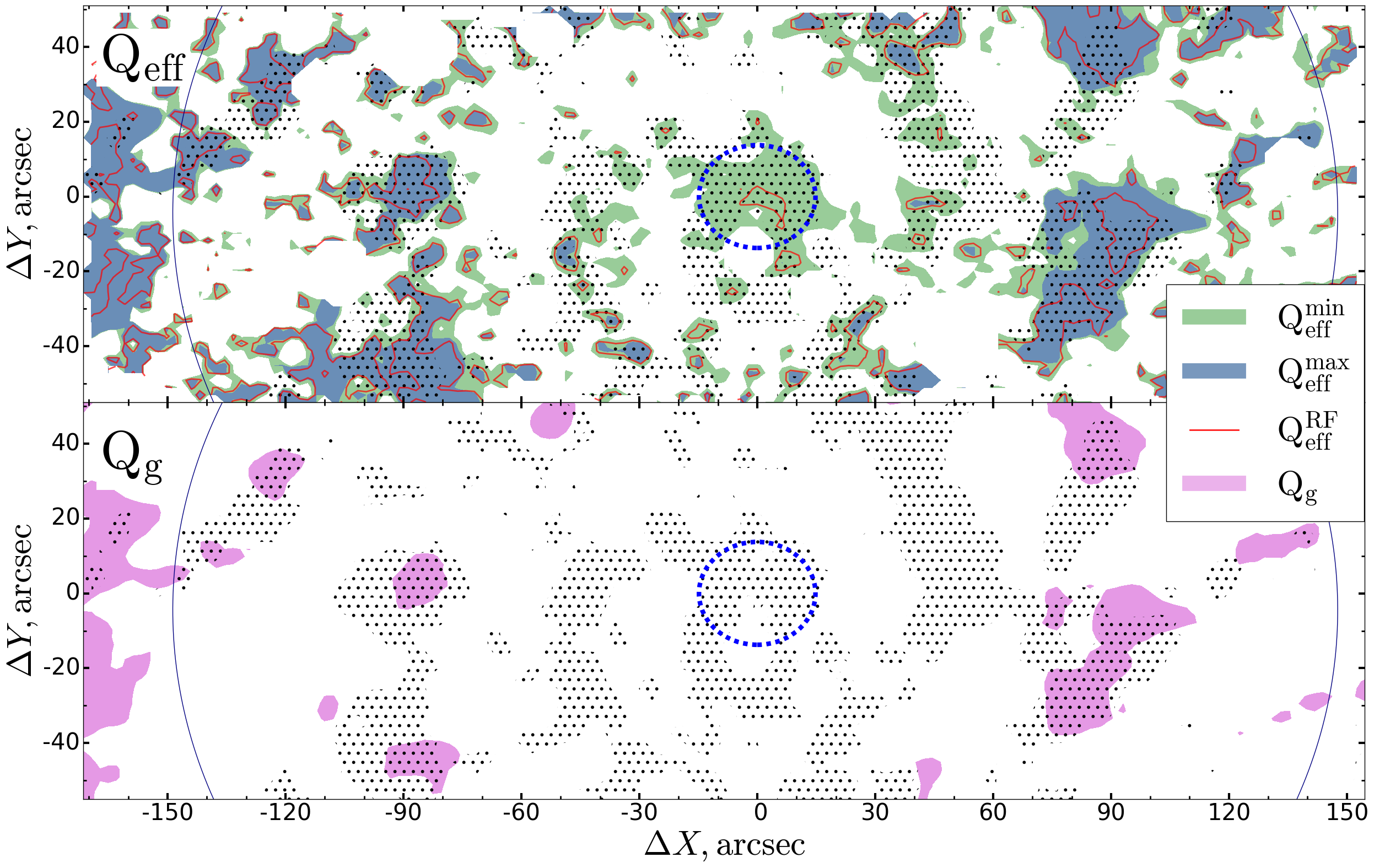}
\caption{Levels of the best coincidence between star formation and gravitational instability. Upper plot shows data for two-component models and bottom for gas only. Color filled areas and red contours correspond to the level $Q\le 3$ of the appropriate model. Hatched regions correspond to $\Sigma_\mathrm{SFR}> 0.007 \, M_{\sun}$\,yr$^{-1}$\,kpc$^{-2}$. Bulge effective radius is marked by the dashed blue line and the area $R \le 140\arcsec$ is enclosed by the blue solid line.}
\label{fig:fig2}
\end{figure*}
%%%%%%%%%%%%%%%%%%%%%%%%%%%%%%%%%%%%%%%%%%%%%%%%%%%%%%%%%%%%%%%%%%%%%

It is interesting to see how well the different instability levels can be associated with star forming regions. This question addresses two threshold values $Q_\mathrm{lim}$ and $\Sigma_\mathrm{SFR}^\mathrm{lim}$. The best matching between gravitationally unstable $Q < Q_\mathrm{lim}$ regions and noticeable star formation areas $\Sigma_\mathrm{SFR} > \Sigma_\mathrm{SFR}^\mathrm{lim}$ is searched for. Thereafter pixels with SFR density lower than threshold will be called ``without signal'' and ``with signal'' otherwise. Pixels where $Q < Q_\mathrm{lim}$ will be mentioned as ``with predicted signal'' and ``without predicted signal'' else. The whole problem was treated as a binary classification task where the exact values of $Q$ or SFR in pixels are not important, but only agreement between signal presence or absence and its correct prediction.

A commonly accepted criteria for coincidence like $\chi^2$ are not appropriate, because high $Q_\mathrm{lim}$ and low $\Sigma_\mathrm{SFR}^\mathrm{lim}$ along with low $Q_\mathrm{lim}$ and large $\Sigma_\mathrm{SFR}^\mathrm{lim}$ show better matching than intermediate values. This point illustrated on the central subplot in Fig.~\ref{fig:match}, where solid lines represent ratio of correctly founded pixels for $Q_\mathrm{lim}=3$ case. This is so-called accuracy metric in machine learning field. It is equal to number of matches between observations and model predictions (i.e. number of pixels ``with signal'' and  ``with predicted signal'' or ``without signal'' and  ``without predicted signal'' accordingly) divided by number of pixels. Accuracy by definition can't be larger than unity.  From central subplot in Fig.~\ref{fig:match} it is seems like $Q_\mathrm{g}$ model explains data better than two-component models for some $\Sigma_\mathrm{SFR}^\mathrm{lim}$ thresholds. However, as mentioned above, this is not completely true, because areas with and without star formation are highly imbalanced for some thresholds values and one can actually get lower total error if predicts no star  formation at all. This can be also seen from additional dashed lines on the same figure, which show how many star-forming regions predicted well. This line represent so-called recall, which equals to number of correctly predicted pixels with signal divided by total number of pixels with signal. Obviously $Q_\mathrm{eff}^\mathrm{min}$ model makes correct signal prediction more often than other models. Note that both types of lines converges on the left border, because there is no pixels without signal anymore and accuracy become the same as recall.

Thus the problem of finding best match is not straightforward and need to be balanced somehow for two types of errors: predict nothing where signal exists and does not predict any where it actually is. One need to find limits which lower both mentioned types of errors simultaneously. For this reason a well known in machine learning approach was used instead. These limits are searched in order to maximize so-called $F_1$-score, which represents balanced metric and equals to harmonic mean between precision (ratio of correctly predicted pixels with signal to total number of pixels with predicted signal) and recall for predicted and observed classes \citep{Rijsbergen1979}. An additional limitation is that a correction for nonaxisymmetric perturbations can not lead to the $Q$ threshold larger than 3 (see e.g. \citealp{Marchuk_Sotnikova2018} and references therein). 

Measured $F_1$ scores for different $Q_\mathrm{eff}^\mathrm{min}$ levels are shown in the left plot in Fig.~\ref{fig:match}. The higher score is better. It can be seen from the plot that increasing $Q_\mathrm{lim}$ allow to find higher score. As possible $Q_\mathrm{lim}$ values are limited by upper value one should search the best score for $Q_\mathrm{lim}=3$ case. Corresponding $F_1$ scores for all models are shown on the right plot in Fig.~\ref{fig:match}. As expected, one-fluid model is less precise than two-component and $Q_\mathrm{eff}^\mathrm{min}$ shows higher scores than $Q_\mathrm{eff}^\mathrm{max}$. Last mentioned model demonstrates maximum for $\log \Sigma_{\mathrm{SFR}} \approx -2.25$. Better model $Q_\mathrm{eff}^\mathrm{min}$ did not reveal clear maximum score, but become flat at level $F_1\approx 0.56$ after $\log \Sigma_{\mathrm{SFR}} \approx -2.5$, which means that model does not become better for lower $\Sigma_\mathrm{SFR}^\mathrm{lim}$ thresholds. Taking into account that solid accuracy line for $Q_\mathrm{eff}^\mathrm{min}$ model on the left plot shows maximum near $\log \Sigma_{\mathrm{SFR}} \approx -2.0$, it must be concluded that best searched match should lies between -2.5 and -2.0 boundaries.

Hence the best obtained fit is for $Q_\mathrm{eff}^\mathrm{min}$ model with $Q_\mathrm{lim} = 3$ and $\Sigma_\mathrm{SFR}^\mathrm{lim}=0.007\, M_{\sun}$\,yr$^{-1}$\,kpc$^{-2}$, which is the average value between boundaries above. Star formation contours in all map figures (Figs.~\ref{fig:fig1},\ref{fig:lambda},\ref{fig:depl},\ref{fig:fig2}) in this work are also not arbitrary and represent this best fit. Corresponding areas for best match thresholds are shown in Fig.~\ref{fig:fig2}. Notice a good agreement between the model and observations even inside the bulge effective radius. As before, the gaseous disc alone can explain star formation only in outer regions even for $Q_\mathrm{g} \approx 3$. Note also that unstable regions according to $Q_\mathrm{eff}^\mathrm{max}$ model lie inside those for $Q_\mathrm{eff}^\mathrm{min}$ model and $Q_\mathrm{RF}^\mathrm{min}$ model is similar to $Q_\mathrm{eff}^\mathrm{max}$ one as was mentioned above. Two-component model of gravitational instability $Q_\mathrm{eff}^\mathrm{min}$ can correctly explain selected regions of star formation for vast radii range. An implicit evidence that founded match is reasonable also come from depletion time map agreement.

\section{Summary}

In this work large number of two-dimensional data were collected for grand design galaxy NGC~628 and the connection between two-component gravitational instability and large-scale star formation was analyzed. An additional support for such relationship came from recent works about self-shielding \citep{Orr2017,Stark2018} and from idea of global Toomre regime \citep{Krumholz2012}. A small number of assumptions were used and the performed analysis is mainly similar to that in previous work \citet{Marchuk_Sotnikova2018}, but adjusted for maps. Various tests of obtained results were performed including azimuthal averaging and accounting for disc thickness effect. Some star formation relations were testified. The best explainable $\Sigma_\mathrm{SFR}^\mathrm{lim}$ threshold was found.

In summary, next results were obtained:
\begin{enumerate}
	\item the two-component gravitational instability model for NGC~628  shows a good agreement with observable large-scale star formation; 
	\item previous result stays also true if the model is accounted for $X_\mathrm{CO}$ gradients or disc thickness effect; 
	\item the parameter $Q_\mathrm{g}$ for gas-only model can explain star formation only in distant regions and thus the destabilizing effect of a stellar component can not be neglected; 
	\item an azimuthal averaging procedure make disc more stable and can not predict the location of star forming regions;
	\item the relationship between disc instability and star formation is perhaps tighter than currently believed at least for one model;
	\item the largest area of star formation in NGC~628 which can be explained by gravitational instability is determined by the condition $\Sigma_\mathrm{SFR}^\mathrm{lim}>0.007\, M_{\sun}$\,yr$^{-1}$\,kpc$^{-2}$.
\end{enumerate}
%%%%%%%%%%%%%%%%%%%%%%%%%%%%%%%%%%%%%%%%%%%%%%%%%%%%%%%%%%%%%%%%%%%%%%
\section*{Acknowledgements}
%%%%%%%%%%%%%%%%%%%%%%%%%%%%%%%%%%%%%%%%%%%%%%%%%%%%%%%%%%%%%%%%%%%%%

Author thanks Alessandro Romeo for his great review and highly appreciate the comments and suggestions that significantly contributed to improving the quality of the article. Author is grateful to Guillermo Blanc for providing VENGA data and acknowledge the hard work of the THINGS, HERACLES, S$^4$G and VENGA teams and thank them for making their data publicly available. 

This research makes use of the NASA/IPAC Extragalactic
Database (NED) which is operated by the Jet Propulsion Laboratory, California Institute of Technology, under contract with the National Aeronautics and Space Administration and the LEDA database (\href{url}{http://leda.univ-lyon1.fr}).
%%%%%%%%%%%%%%%%%%%%%%%%%%%%%%%%%%%%%%%%%%%%%%%%%%%%%%%%%%%%%%%%%%%%%%

%%%%%%%%%%%%%%%%%%%%%%%%%%%%%%%%%%%%%%%%%%%%%%%%%%%%%%%%%%%%%%%%%%%%%%
\bibliographystyle{mnras}
\bibliography{art}
%%%%%%%%%%%%%%%%%%%%%%%%%%%%%%%%%%%%%%%%%%%%%%%%%%%%%%%%%%%%%%%%%%%%%%
\label{lastpage}
%%%%%%%%%%%%%%%%%%%%%%%%%%%%%%%%%%%%%%%%%%%%%%%%%%%%%%%%%%%%%%%%%%%%%%
\end{document}